\documentclass[showpacs,amsmath,amssymb,aps]{revtex4} 
\usepackage{graphicx}
\usepackage{epstopdf}
\usepackage{amsmath}
\usepackage{bm}

\begin{document}

\title{Higgs-like mechanism for spontaneous spacetime symmetry breaking}

\author{Kimihide Nishimura}
\email{kimihiden@dune.ocn.ne.jp}
\affiliation{Nihon-Uniform, 1-4-1 Juso-Motoimazato, Yodogawa-ku, Osaka 532-0028, Japan}

\date{\today}

\begin{abstract}
The study of spontaneous breakdown of spacetime symmetries leads to the discovery of another type of Higgs mechanism operating in a chiral SU(2) model. Some of the Nambu-Goldstone vector mesons emergent from simultaneous violations of gauge and Lorentz symmetries are, in this case, absorbed by a left-handed doublet and endow one of the fermions with a right-handed state, while another part becomes emergent as photons. 
Accordingly, this mechanism allows a chiral fermion to acquire a mass, and it may enable the emergent theory to reproduce the electromagnetism equivalent to the QED sector in the standard theory. It is also mentioned that the ``fermion-boson puzzle" known in the presence of a 't Hooft--Polyakov monopole does not exist in our theory. 
\end{abstract}

\maketitle
\centerline{ }
\centerline{ \em Published in Physical Review D $\bm{92}$, 076010 (2015); doi:10.1103/PhysRevD.92.076010 }

\section{Introduction}
We understand under the name of the Higgs mechanism \cite{Higgs1,Higgs2,Englert} that the Nambu-Goldstone bosons \cite{NJ,Goldstone,GSW} emergent from spontaneous symmetry breaking are absorbed by gauge bosons, and the gauge bosons in turn acquire longitudinal polarization states and become massive. 
We report here that if, in addition to gauge symmetries, spacetime symmetries are also spontaneously broken, another type of Higgs mechanism can be operative. The new mechanism generates from a left-handed doublet and Nambu-Goldstone vector mesons a right-handed isosinglet in the sense of the standard theory, and permits a chiral fermion to acquire a Lorentz-invariant mass.

It has been already shown in another paper \cite{KN1} that when SU(2) gauge bosons become massive, one of the fermions in a left-handed doublet coupled with massive gauge bosons can acquire mass due to spontaneous spacetime symmetry breaking. 
If the effective theory is Lorentz invariant, the above result implies some kind of dynamical generation of a right-handed fermion from left-handed ones, since we can always find a Lorentz frame which sees a left-handed massive fermion as a right-handed one.

This paper clarifies the mechanism of this phenomenon, and it shows that a right-handed fermion is in fact generated from a left-handed doublet by absorbing the Nambu-Goldstone mesons resulting from spontaneous spacetime symmetry breaking.
Since this mechanism resembles the Higgs mechanism mentioned above, we may call it a  ``quasi-Higgs" mechanism.

The quasi-Higgs mechanism does not absorb all the Nambu-Goldstone mesons, but part of them emerge as ``photons".
Though the emergence of photons by Lorentz violation has been repeatedly discussed, our photons are of a special kind, which are generated from a model without U(1) currents, and automatically decouple from emergent neutrinos.
Accordingly, the evidence presented in this paper all suggests that the effective theory of our model may be equivalent to the leptonic sector of the standard electroweak theory.

We consider in this paper the system which consists of a chiral SU(2) doublet $\varphi$, SU(2) Yang-Mills gauge fields $\bm{Y}^{\mu}$, and a Higgs doublet $\Phi$.
 If the Higgs doublet is absent, and spontaneous spacetime-gauge symmetry breaking does not occur, the model potentially has the Witten anomaly \cite{Witten}, which would make our model inconsistent. However, the argument on the Witten anomaly is essentially based on a Z$_2$ symmetry in the configuration space of SU(2) Yang-Mills fields for local gauge invariant models with odd number of chiral doublets. 
 Therefore, it is questionable whether this anomaly persists even when the physical vacuum breaks these symmetries. In any case, as the arguments in this paper never encounter the situation in which the Witten anomaly becomes an issue, we will not argue any further about this anomaly in the context of the present theory.

The differences between our system and the leptonic sector of the standard electroweak theory are the absence of the right-handed fermion (electron), a U(1) gauge boson (photon), the Yukawa coupling with Higgs fields to make the fermion mass term, and the Yukawa coupling of the fermions with U(1) gauge field.

What this paper would like to show is that spontaneous Lorentz  symmetry breaking generates dynamically the left- and right-handed massive fermions (electrons) with the same mass, a left-handed massless fermion (neutrino), and the massless vector bosons (photons) coupling to massive fermions (electrons), but not to a massless fermion (neutrino). 
Then, the confirmation of these results will conclude that the spacetime symmetry breaking makes our model almost equivalent to the leptonic sector of the standard electroweak theory.  The only difference will be that the fermion mass originates from a new dynamics, not from the Yukawa coupling with Higgs fields.

It should be remarked in advance, however, that the demonstrations presented in this paper to show the equivalence between the two theories are still  partly inadequate.
The inadequacies mainly lie in that we have not yet succeeded in rewriting the effective action in terms of the Dirac spinors corresponding to emergent neutrinos and electrons, nor in  translating the SU(2) symmetry in our model into that of the standard theory. 
These inadequacies are, however, compensated by another arguments, and they will not essentially affect the conclusions of this paper.

The spontaneous spacetime symmetry breaking in our model is caused by the vacuum expectation values of gauge potentials, which might invoke a phenomenon similar to that affecting  
the alteration of statistics in the presence of a 't Hooft--Polyakov monopole \cite{HGT,JR,ASG}.
We show that, though the isospin actually incorporates with the spin, the effect in the present case is not to transmute fermions into bosons, but rather to provide a chiral fermion with an additional spin state. Therefore, the problem with statistics does not exist in our model.
%%%%%%%%%%%%%%%%%%%%%%%%%%%%%%%%%
\section{spontaneous Spacetime Symmetry Breaking\label{SSSB}}

After spontaneous ``gauge" symmetry breaking, we approximate our system with the following Lagrangian:
\begin{equation}
{\cal L}=-\frac{1}{4}\bm{Y}^{\mu\nu}\cdot\bm{Y}_{\mu\nu}+\frac{1}{2}m^2_Y\bm{Y}^{\mu}\cdot\bm{Y}_{\mu}
+\varphi^\dagger\bar{\sigma}^\mu i\partial_\mu\varphi-\bm{j}^\mu\cdot\bm{Y}_\mu,
\label{Lag}
\end{equation}
\begin{equation}
\begin{array}{ccc}\bm{j}^\mu=g\varphi^\dagger\bar{\sigma}^\mu\displaystyle\frac{\bm{\rho}}{2}\varphi,&\varphi=\left(
\begin{array}{c}\varphi_+\\ \varphi_-\end{array}\right),&\bar{\sigma}^\mu=(1,-\bm{\sigma}),
\end{array}
\label{SU(2)current}
\end{equation}
where $\varphi_\pm$  are left-handed Weyl spinors, $\bm{\rho}/2$ are SU(2) generators, and $g$ is a coupling constant. 
The gauge boson mass term has been  provided by the vacuum expectation values of the Higgs fields, and the remaining Higgs boson terms have been eliminated.
It is also assumed for simplicity that the Lagrangian (\ref{Lag}) has only a global SU(2) symmetry, and therefore $\bm{Y}^{\mu\nu}$ contains no non-Abelian part. 
The model with local SU(2) symmetry was discussed in the previous paper \cite{KN1} on the spacetime symmetry breaking. 

The occurrence of Lorentz symmetry violation in the simplified model (\ref{Lag}) is demonstrated as follows.
We assume first that $\bm{Y}^\mu$ and $\bm{j}^\mu$ have  constant vacuum expectation values $\bm{\zeta}_0^\mu$ and $\bm{J}_0^\mu$, respectively, and decompose the original fields as
\begin{equation}
\left\{\begin{array}{lll}
\bm{Y}^\mu=\bm{\zeta}_0^\mu+\bm{Z}^\mu, 
&\bm{\zeta}_0^\mu=\langle \bm{Y}^\mu\rangle,&\langle\bm{Z}^\mu\rangle=0,\\
\bm{j}^\mu=\bm{J}_0^\mu+:\bm{j}^\mu:, 
&\bm{J}_0^\mu=\langle \bm{j}^\mu\rangle,
&\langle:\bm{j}^\mu:\rangle=0.
\end{array}\right.
\label{DecSU(2)}
\end{equation}
Then Lagrangian (\ref{Lag}) is rewritten as
\begin{eqnarray}
{\cal L}&=&-\frac{1}{4}\bm{Z}^{\mu\nu}\cdot\bm{Z}_{\mu\nu}+\frac{m_Y^2}{2}\bm{Z}^\mu\cdot\bm{Z}_\mu
\nonumber\\
&&+\bm{Z}_\mu\cdot\left(m_Y^2\bm{\zeta}_0^\mu-\bm{J}_0^\mu\right)
+\frac{m_Y^2}{2}\bm{\zeta}_0^\mu\cdot\bm{\zeta}_{0\mu}
\nonumber\\
&&+\varphi^\dagger\bar{D}\varphi-:\bm{j}^\mu:\cdot\bm{Z}_\mu,
\label{Lag_R}
\end{eqnarray}
where
\begin{equation}
\begin{array}{cc}
\bar{D}=i\bar{\sigma}^\mu\partial_\mu-\bar{M},&\bar{M}=g\bar{\sigma}^\mu \displaystyle\frac{\bm{\rho}}{2}\cdot\bm{\zeta}_{0\mu}.
\end{array}
\label{DZ}
\end{equation}
The stability of the vacuum under the $\bm{Z}$-boson  production requires that the linear term with respect to $\bm{Z}^\mu$ should vanish: 
\begin{equation}
\bm{J}_0^\mu=m_Y^2\bm{\zeta}_0^\mu.
\label{SSSBCSU(2)}
\end{equation}
The value of $\bm{J}_0^\mu$ can be calculated as
\begin{equation}
\bm{J}_0^\mu=-g{\rm Tr}\bar{\sigma}^\mu\displaystyle\frac{\bm{\rho}}{2} S(0),
\label{DefJ^mu}
\end{equation}  
with the help of the fermion propagator 
\begin{equation}
S(x)=\displaystyle\int\frac{d^4p}{(2\pi)^4}
\frac{i}{\bar{\sigma}\cdot p-\bar{M}}e^{-ip\cdot x}.
\label{FP}
\end{equation}
The contour of $p_0$ integration in (\ref{FP}) is specified in Appendix \ref{LIEDI}. 
Then, Eq. (\ref{SSSBCSU(2)}) gives the self-consistency condition with respect to $\bm{\zeta}_0^\mu$ for the occurrence of spontaneous Lorentz symmetry breaking.
Incidentally, we can show that (\ref{SSSBCSU(2)}) is equivalent to the extremum condition for the potential density $v(\bm{Y}^\mu)$ of $\bm{Y}^\mu$.
In fact, writing for each $a$, $Y_a^\mu=e_a^\mu Y_a$ and $j_a^\mu=e_a^\mu j_a$ with constant unit four-vectors $e_a^\mu$: $e_a\cdot e_a=\pm1$, we find
\begin{equation}
v(\bm{Y}^\mu)=-\frac{m_Y^2}{2}\bm{Y}^\mu\cdot\bm{Y}_\mu+\bm{j}^\mu\cdot\bm{Y}_\mu
=\sum_a-e_a\cdot e_a\left(\frac{m_Y^2}{2}Y_a^2-j_aY_a\right),
\label{PDforY}
\end{equation} 
which attains the minimum at $Y_a=j_a/m_Y^2$ if $e_a\cdot e_a=-1$ for every $a$, but it turns to the maximum if $e_a\cdot e_a=1$  for every $a$. 
Accordingly, in order to occur spontaneous spacetime symmetry breaking, it is necessary to satisfy not only Eq. (\ref{SSSBCSU(2)}), but also $\bm{\zeta}_0^\mu$ to be spacelike vectors. 

We next show that $\bar{M}=\displaystyle\frac{m}{2}\bm{\rho}\cdot\bm{\sigma}$, which corresponds to spacelike $\bm{\zeta}_0^\mu$: $(\zeta_0)_a^0=0$, $(\zeta_0)_a^i=\delta_a^im/g$, can result from spontaneous spacetime symmetry violation. Actually as calculated in Appendix \ref{LIEDI}, we obtain   
\begin{eqnarray}
\bm{J}_0^\mu=g^2\Gamma\bm{\zeta}_0^\mu&,&
\Gamma=\displaystyle\frac{k_1}{2}-\frac{2m^2}{3}\left(\frac{5}{96\pi^2}-K_2(m)\right),\label{Gamma}\\
k_1=\displaystyle\int\frac{d^4p}{(2\pi)^4}\frac{i}{p^2+i\epsilon}&,&K_2(m)=\displaystyle\int\frac{d^4p}{(2\pi)^4}\frac{i}{(p^2-m^2+i\epsilon)^2}.
\end{eqnarray}
Then the self-consistency condition (\ref{SSSBCSU(2)}) reduces to $g^2\Gamma/m_Y^2=1$, which gives 
\begin{equation}
m^2=\frac{3}{2}\cdot\frac{g^2k_1/2-m_Y^2}{5/(96\pi^2)-K_2(m)}.
\label{SCC}
\end{equation}
Since $K_2(m)=-\sum_{\bm{p}}\omega^{-3}/4<0$, and its $m$ dependence is weak (logarithmic), Lorentz symmetry breaks if  $g^2k_1/2>m_Y^2$. In this case, the dispersion relations of an emergent fermion doublet with 4-momentum $p^\mu=(p_0, \bm{p})$ obey
\begin{equation}
\vert\bar{\sigma}\cdot p-\bar{M}\vert=\left(p\cdot p-\frac{m^2}{4}\right)^2
-m^2(p_0-\frac{m}{2})^2=0,
\label{DispersionLaws}
\end{equation}
from which we obtain the following solutions and corresponding wave functions \cite{KN1}:
\begin{equation}
\begin{array}{lll}
p_0=|\bm{p}|+m/2&:&\varphi_{1\bm{p}}=\chi_{\bm{p}L}\varphi_{\bm{p}L},\\
p_0=\omega-m/2&:&\varphi_{2\bm{p}}=\lambda_{+}\chi_{\bm{p}R}\varphi_{\bm{p}L}+\lambda_{-}\chi_{\bm{p}L}\varphi_{\bm{p}R},
\\
p_0=-|\bm{p}|+m/2&:&\varphi_{3\bm{p}}=\chi_{\bm{p}R}\varphi_{\bm{p}R},
\\
p_0=-\omega-m/2&:&\varphi_{4\bm{p}}=\lambda_{+}\chi_{\bm{p}L}\varphi_{\bm{p}R}-\lambda_{-}\chi_{\bm{p}R}\varphi_{\bm{p}L}.
\end{array}
\label{DR&WF}
\end{equation}
\begin{equation}
\begin{array}{cc}
\omega=\sqrt{\bm{p}^2+m^2},&
\lambda_\pm=\displaystyle\frac{1}{2}\left[\sqrt{1+\frac{m}{\omega}}\pm\sqrt{1-\frac{m}{\omega}}\right],
\end{array}
\end{equation}
\begin{equation}
\begin{array}{ll}
\varphi_{\bm{p}R}=\left(
\begin{array}{c}
\cos(\theta/2)\\
\sin(\theta/2)e^{i\phi}
\end{array}
\right),&
\varphi_{\bm{p}L}=\left(\begin{array}{c}
-\sin(\theta/2)e^{-i\phi}\\
\cos(\theta/2)
\end{array}
\right),
\\
\chi_{\bm{p}R}=\left[
\begin{array}{c}
\bm{1}\cos(\theta/2)\\
\bm{1}\sin(\theta/2)e^{i\phi}
\end{array}
\right],&
\chi_{\bm{p}L}=\left[
\begin{array}{c}
-\bm{1}\sin(\theta/2)e^{-i\phi}\\
\bm{1}\cos(\theta/2)
\end{array}
\right],
\end{array}
\label{PX}
\end{equation}
where $(\theta, \phi)$ are the polar coordinates of 3-momentum $\bm{p}$.
$\varphi_{\bm{p}R}$ and $\varphi_{\bm{p}L}$ are 2-spinors in  helicity eigenstates satisfying $h\varphi_{\bm{p}R}=\varphi_{\bm{p}R}$ and $h\varphi_{\bm{p}L}=-\varphi_{\bm{p}L}$ with $h=\bm{\sigma}\cdot\bm{p}/\vert\bm{p}\vert$, while $\chi_{\bm{p}R}$ and $\chi_{\bm{p}L}$ are SU(2) doublets, in which $\bm{1}$ denotes a 2$\times$2 unit matrix.
We also notice that the dispersion relation 
(\ref{DispersionLaws}) is invariant under a SU(2) transformation of the mass matrix $\bar{M}$:
\begin{equation}
|\bar{\sigma}\cdot p-\bar{M}'|=|U(\bar{\sigma}\cdot p-\bar{M})U^{-1}|=|\bar{\sigma}\cdot p-\bar{M}|=0,
\label{SU(2)equivalence}
\end{equation}
which shows that the dispersion relations in (\ref{DR&WF}) ---and therefore the properties of emergent fermions---  are invariant under SU(2) transformations.

If we identify $m$ with the electron mass $m_e$,
the first and second solutions in (\ref{DR&WF}) become equivalent to the dispersion relations for neutrinos and electrons, respectively, since additive constant terms $\pm m/2$ could be absorbed by local phase transformations for their effective Dirac fields \cite{CK1,CK2}. On the other hand, the third and fourth solutions correspond to negative energy states, if additive terms are similarly removed. Accordingly, if the hole interpretation is applied, they will be equivalent to the dispersion relations for antineutrinos and positrons, respectively. 

The hole theory assumes for the vacuum that all of the negative energy states are already occupied, and the only thing that we can do is to annihilate the negative energy particles. The absence of a particle with 4-momentum $p^\mu$ is interpreted as a hole with 4-momentum $-p^\mu$. Similarly, the annihilation operator of a negative energy particle with momentum $\bm{p}$ is then reinterpreted as the creation operator of an antiparticle with momentum $-\bm{p}$. We apply this hole interpretation to  
the third and fourth solutions in (\ref{DR&WF}), which implies that we have generalized the hole theory to make it applicable to the  particles with potential energies. The generalized hole theory  applies not to negative energy particles, but to particles corresponding to the negative branch of the square root in the dispersion relation. 
In the above sense, the hole of the third particle in (\ref{DR&WF}) has energy 
$|\bm{p}|-m/2$, while that of the fourth particle has energy 
$\omega+m/2$.
Similarly, under the momentum reflection $\bm{p}\rightarrow-\bm{p}$, the wave functions are replaced according to 
$\varphi_{\bm{p}R}\leftrightarrow\varphi_{\bm{p}L}$ and  
$\chi_{\bm{p}R}\leftrightarrow \chi_{\bm{p}L}$, except for  inessential phase factors.

From above considerations, we can construct the interaction representation of the chiral doublet operator $\varphi$ in terms of the creation and annihilation operators for emergent fermions: 
\begin{equation}
\varphi(x)=V^{-1/2}\sum_{\bm{p}}\left[
\begin{array}{r}\nu_{\bm{p}}\varphi_{\nu\bm{p}} e^{-ip_\nu\cdot x}+\bar{\nu}_{\bm{p}}^\dagger\varphi_{\bar{\nu}\bm{p}} e^{ip_{\bar{\nu}}\cdot x}\\
+e_{\bm{p}}\varphi_{e\bm{p}} e^{-ip_e\cdot x}+\bar{e}_{\bm{p}}^\dagger\varphi_{\bar{e}\bm{p}} e^{ip_{\bar{e}}\cdot x}\end{array}
\right],
\label{Operator_varphi}
\end{equation}
\begin{equation}
\begin{array}{cc}
\left\{\begin{array}{c}
p^\mu_\nu=(|\bm{p}|+\displaystyle\frac{m}{2},\bm{p}),\\
\\
p^\mu_{\bar{\nu}}=(|\bm{p}|-\displaystyle\frac{m}{2},\bm{p}),
\end{array}\right.
&
\left\{\begin{array}{c}
p^\mu_e=(\omega-\displaystyle\frac{m}{2},\bm{p}),\\
\\
p^\mu_{\bar{e}}=(\omega+\displaystyle\frac{m}{2},\bm{p}),
\end{array}\right.
\end{array}
\label{DR}
\end{equation}
\begin{equation}
\left\{\begin{array}{l}
\varphi_{\nu\bm{p}}=\varphi_{\bar{\nu}\bm{p}}=\chi_{\bm{p}L}\varphi_{\bm{p}L},\\
\varphi_{e\bm{p}}=\lambda_+\chi_{\bm{p}R}\varphi_{\bm{p}L}+\lambda_-\chi_{\bm{p}L}\varphi_{\bm{p}R},\\
\varphi_{\bar{e}\bm{p}}=\lambda_+\chi_{\bm{p}R}\varphi_{\bm{p}L}-\lambda_-\chi_{\bm{p}L}\varphi_{\bm{p}R}.\\
\end{array}\right.
\label{WaveFunctions}
\end{equation}
We regard $\nu_{\bm{p}}$ and $\bar{\nu}_{\bm{p}}$ as annihilation operators of the emergent neutrinos and antineutrinos with momentum $\bm{p}$, respectively, satisfying the ordinary anticommutation relations: $\{\nu_{\bm{p}}, \nu_{\bm{p}'}^\dagger\}=\delta_{\bm{p},\bm{p}'}$ etc., while $e_{\bm{p}}$ and $\bar{e}_{\bm{p}}$ are those for the emergent electrons and positrons. 
The field operator $\varphi(x)$ defined by 
(\ref{Operator_varphi}) indeed satisfies the ordinary second quantization conditions.

Besides extra potential terms: $\pm m/2$ in the dispersion relations,  the other difference between emergent leptons and real leptons is that an emergent electron has only one spin state, just as a neutrino. 
However, the emergent electron comes to have two spin states owing to the ``Higgs-like" mechanism explained in Sec.\ref{ERHF}.
%%%%%%%%%%%%%%%%%%%%%%%%%%%%%%%%
\section{Emergent photons\label{EP}}
We here turn to the proof of the emergence of Nambu-Goldstone photons. Dynamical photons by Lorentz symmetry violation have been recurrently discussed since 1960 \cite{Bjorken, Guralnik,Eguchi}. For several reasons, however, those photons will not have direct relevance to the Nambu-Goldstone photons emergent from the present model, nor to the photons in the standard model. 
One of the reasons is that most old papers argue dynamical photons based on  models with four-fermion interactions, 
which have already been obsolete since the establishment of electroweak gauge theory.  In addition, the old models include U(1) currents from the beginning, while our model contains SU(2) currents only.

The second concerns the assumption of the vacuum expectation value $J^\mu$ of U(1) current $j^\mu$. For example, Bjorken \cite{Bjorken} assumes that $J^\mu$ is a timelike vector, and the vacuum has charge asymmetry. Guralnik \cite{Guralnik} also  discusses the case with a lightlike vector: $J^2=0$. As we saw  in Sec.~\ref{SSSB}, their assumptions contradict the requirement---$J^2<0$ for the spontaneous space-time symmetry breaking---, even disregarding the difference between  models.  

The third reason concerns the Eguchi's derivation of Bjorken photons. Eguchi \cite{Eguchi} mentions in his paper which proves the equivalence of a four-fermion interaction model with QED that the Bjorken photons actually emerge by renormalization, and the condition of the vanishing of the photon mass is the same as that of Lorentz violation.
Nevertheless, he rejects the view of Bjorken photons as Nambu-Goldstone mesons emergent from spontaneous Lorentz violation. 

The four-fermion interaction made of a bilinear form of  currents can be regarded as a low energy approximation of interactions mediated by massive vector bosons. This interpretation allows us to regard Eguchi's arguments and results as approximately applicable to the present case. However, if we follow this view,  Eguchi's proof of equivalence can be reinterpreted such that the gauge bosons which acquire mass by the Higgs mechanism return massless again to become Bjorken photons due to the fermion loop corrections. 
We come, accordingly, to the conclusion that Eguchi's renormalization theory would imply the ``anti-Higgs" mechanism, rather than the dynamical generation of photons, and that the four-fermion interaction could not serve as a low energy approximation of weak interactions. In this respect, Eguchi's disapproval of viewing Bjorken photons as Nambu-Goldstone bosons seems fair, since spontaneous spacetime symmetry breaking should generate Nambu-Goldstone photons not by depriving massive gauge bosons of their mass but by generating new massless vector bosons, besides massive vector bosons. 

Though Eguchi's method of deriving Bjorken photons by fermion loop corrections cannot be directly applicable for obtaining Nambu-Goldstone photons, we show in the following that a slight improvement will make his method serve our purpose.   

First of all, we remark that due to the invariance of the model under Lorentz and global SU(2) transformations, if some constant vectors $\bm{\zeta}_0^\mu$ and $\bm{J}_0^\mu$ satisfy the condition (\ref{SSSBCSU(2)}), then $\bm{\zeta}^\mu$ and $\bm{J}^\mu$, obtained by Lorentz and gauge transformations
\begin{equation}
\begin{array}{cc}
\bm{\zeta}^\mu=\Lambda^\mu{}_\nu O\bm{\zeta}_0^\nu,
&\bm{J}^\mu=\Lambda^\mu{}_\nu O\bm{J}_0^\nu,
\end{array}
\end{equation}
also satisfy (\ref{SSSBCSU(2)}), where $O$ denotes an SO(3) matrix, and $\Lambda^\mu{}_\nu$ is a Lorentz transformation matrix.
Since the potential density (\ref{PDforY}) is also the same for $\bm{\zeta}^\mu$ and $\bm{J}^\mu$, an infinitesimal fluctuation $\delta\bm{\zeta}^\mu=\bm{\zeta}^\mu-\bm{\zeta}_0^\mu$ with $\delta\bm{J}^\mu=\bm{J}^\mu-\bm{J}_0^\mu$ does not alter the vacuum energy, which in turn guarantees that the quanta corresponding to $\delta\bm{\zeta}^\mu$ are massless.

We first promote the generalized vacuum expectation values  
$\bm{\zeta}^\mu$ and $\bm{J}^\mu$ introduced above to spacetime dependent $\bm{\zeta}^\mu(x)$ and $\bm{J}^\mu(x)$ in order to account for Nambu-Goldstone fluctuations  
as background fields, 
and we separate the constant part $\bm{\zeta}_0^\mu$ from $\bm{\zeta}^\mu$ as 
\begin{eqnarray}
\bm{\zeta}_0^\mu\rightarrow
\bm{\zeta}^\mu(x)=\bm{\zeta}_0^\mu+\delta\bm{\zeta}^\mu(x)&,&
\bm{\zeta}_0^\mu=\int d^4x\bm{\zeta}^\mu(x)/\int d^4x,
\label{DecofNGM}
\end{eqnarray}
while we decompose $\bm{j}^\mu$ as
\begin{equation}
\begin{array}{lll}
\bm{j}^\mu=\bm{J}^\mu+:\bm{j}^\mu:, 
&\bm{J}^\mu=\langle \bm{j}^\mu\rangle,
&\langle:\bm{j}^\mu:\rangle=0.
\end{array}
\label{Dec_R)}
\end{equation}
The introduction of Nambu-Goldstone fields $\bm{\zeta}^\mu(x)$ in this way appears to be equivalent to the decomposition $\bm{Y}^\mu=\bm{\zeta}^\mu+\bm{Z}^\mu$, which may give an impression of $\bm{\zeta}^\mu$ as double-counting, or redundant variables.
However, in contrast to the ordinary spontaneous symmetry breaking triggered by Higgs fields, the Nambu-Goldstone bosons $\bm{\zeta}^\mu$ in the present theory are dynamical mesons in nature, composed of fermions, not part of the triggering fields already present in a model.
Furthermore, in the ladder approximation, the calculation of the dynamical meson propagator cannot be distinguished from the renormalization of gauge boson propagator by the vacuum polarization.
In this view, the double-counting-like appearance of mathematical introduction of the Nambu-Goldstone modes $\bm{\zeta}^\mu(x)$ will not be avoidable.
In the following calculations we distinguish $\bm{\zeta}^\mu$ from $\bm{Z}^\mu$ by treating $\bm{\zeta}^\mu$ simply as parameters to make the vacuum stable under $\bm{Z}^\mu$-boson production, similarly to $\bm{\zeta}_0^\mu$ in Sec.~\ref{SSSB}.
If spontaneous Lorentz violation does not occur, then $\bm{\zeta}_0^\mu=0$, and  we understand $\bm{\zeta}^\mu(x)=0$.

Under these premises, the Lagrangian (\ref{Lag_R}) is rewritten as
\begin{eqnarray}
{\cal L}&=&-\frac{1}{4}\bm{Z}^{\mu\nu}\cdot\bm{Z}_{\mu\nu}+\frac{m_Y^2}{2}\bm{Z}^\mu\cdot\bm{Z}_\mu
\nonumber\\
&&+\bm{Z}_\mu\cdot\left(m_Y^2\bm{\zeta}^\mu-\bm{J}^\mu\right)
+\frac{m_Y^2}{2}\bm{\zeta}^\mu\cdot\bm{\zeta}_\mu
\nonumber\\
&&+\varphi^\dagger\left[ \bar{D}-g\bar{\sigma}^\mu\frac{\bm{\rho}}{2}\cdot\delta\bm{\zeta}^\mu\right]\varphi-:\bm{j}^\mu:\cdot\bm{Z}_\mu,
\label{Lag_NG}
\end{eqnarray}
and the self-consistency equation (\ref{SSSBCSU(2)}) changes to 
\begin{equation}
m_Y^2\bm{\zeta}^\mu-\bm{J}^\mu=0.
\label{SSSBC}
\end{equation}
We next integrate out the fermion part from Lagrangian (\ref{Lag_NG}) by the path integral method. In order for that, we do the replacement 
\begin{equation}
:\bm{j}^\mu:\rightarrow\bm{j}^\mu-\bm{J}^\mu(\delta\bm{Y}),
\label{replacement}
\end{equation}
with a new vacuum expectation value $\bm{J}^\mu(\delta\bm{Y})$. Then, Lagrangian (\ref{Lag_NG}) is rewritten as  
\begin{eqnarray}
{\cal L}&=&-\frac{1}{4}\bm{Z}^{\mu\nu}\cdot\bm{Z}_{\mu\nu}+\frac{m_Y^2}{2}\bm{Z}^\mu\cdot\bm{Z}_\mu
\nonumber\\
&&+\frac{m_Y^2}{2}\bm{\zeta}^\mu\cdot\bm{\zeta}_\mu
\nonumber\\
&&+\varphi^\dagger\left(\bar{D}-g\delta\bar{Y}\right)\varphi+\bm{Z}^\mu\cdot\bm{J}_\mu(\delta\bm{Y}),
\label{Lag_RR}
\end{eqnarray}
where 
\begin{equation}
\delta\bar{Y}=\bar{\sigma}^\mu\frac{\bm{\rho}}{2}
\cdot(\delta\bm{\zeta}_\mu+\bm{Z}_\mu).
\label{DefYbar}
\end{equation}
The last term in (\ref{Lag_RR}) improves Eguchi's method, which enables us to treat the Nambu-Goldstone bosons by  renormalization. If spontaneous spacetime symmetry breaking does not occur, then $\bm{\zeta}^\mu=0$, and the last term disappears consistently from (\ref{Lag_RR}). 
Owing to the replacement (\ref{replacement}), we can regard $\varphi^\dagger(\bar{D}-g\delta\bar{Y})\varphi$ as the free Lagrangian for fermions, which allows us to treat $\delta\bm{Y}^\mu$ as the background field.
The argument $\delta\bm{Y}$ of $\bm{J}^\mu(\delta\bm{Y})$ has reflected this situation. The fermion path integral replaces the fermion part in (\ref{Lag_RR}) with $\Delta{\cal L}(\delta\bm{Y})$ defined by
\begin{equation}
\Delta{\cal L}(\delta\bm{Y})=-i{\rm Tr}\ln(1-g\bar{D}^{-1}\delta\bar{Y}),
\end{equation}
which can be expanded as
\begin{eqnarray}
\Delta{\cal L}(\delta\bm{Y})=-i{\rm Tr}\ln(1-X)
=i{\rm Tr}\left[X+\frac{1}{2}X^2+\frac{1}{3}X^3+\cdots\right]&,&X=g\delta\bar{Y}\bar{D}^{-1}.
\label{DL}
\end{eqnarray}
On the other hand, $\bm{J}^\mu(\delta\bm{Y})$ is expressible as
\begin{equation}
\bm{J}^\mu(\delta\bm{Y})=-i{\rm Tr}g\bar{\sigma}^\mu\frac{\bm{\rho}}{2}(\bar{D}-g\delta\bar{Y})^{-1}
=-i{\rm Tr}\left[g\bar{\sigma}^\mu\frac{\bm{\rho}}{2}\bar{D}^{-1}\left(1+X+X^2+\cdots\right)\right],
\label{J(Y)}
\end{equation}
from which we obtain the relation:
\begin{equation}
\bm{J}^\mu(\delta\bm{Y})=-\frac{\partial}{\partial\delta\bm{\zeta}_\mu}\Delta{\cal L}(\delta\bm{Y}).
\label{J(Y)&DL(Y)}
\end{equation} 
Then, the effective Lagrangian resulting from the integration is given by 
\begin{eqnarray}
{\cal L}_{\rm eff}&=&-\frac{1}{4}\bm{Z}^{\mu\nu}\cdot\bm{Z}_{\mu\nu}+\frac{m_Y^2}{2}\bm{Z}^\mu\cdot\bm{Z}_\mu
+\frac{m_Y^2}{2}\bm{\zeta}^\mu\cdot\bm{\zeta}_\mu
\nonumber\\
&&+\Delta{\cal L}(\delta\bm{Y})-\bm{Z}^\mu\cdot\frac{\partial\Delta{\cal L}(\delta\bm{Y})}{\partial\delta\bm{\zeta}^\mu}.
\label{EffecLSU(2)}
\end{eqnarray}
The fermion loop correction $\Delta{\cal L}(\delta\bm{Y})$ can be estimated as
\begin{eqnarray}
\Delta{\cal L}(\delta\bm{Y})&=&-\bm{J}^\mu_0\cdot\delta \bm{Y}_\mu+\frac{1}{2}\Pi^{\mu\nu}_{ab}(0)\delta Y_{\mu a}\delta Y_{\nu b}+\frac{1}{12}g^2K_2(m)\delta\bm{Y}^{\mu\nu}\cdot\delta\bm{Y}_{\mu\nu}
+\Delta'{\cal L}(\delta\bm{Y}),\label{DLFSU(2)}\\
\Pi^{\mu\nu}_{ab}(0)&=&\int\frac{d^4p}{(2\pi)^4}i{\rm Tr}\left[g\bar{\sigma}^\mu\frac{\rho_a}{2}(\bar{\sigma}\cdot p-\bar{M})^{-1}g\bar{\sigma}^\nu\frac{\rho_b}{2}(\bar{\sigma}\cdot p-\bar{M})^{-1}\right]=-\frac{\partial( J_0)^\mu_a}{\partial (\zeta_0)_{\nu b}}=-m_Y^2g^{\mu\nu}\delta_{ab},
\label{Pi^munu_ab}
\end{eqnarray}
where $\Delta'{\cal L}(\delta\bm{Y})$ is finite. 
Of the equalities in (\ref{Pi^munu_ab}), the first is the definition of $\Pi^{\mu\nu}_{ab}(0)$, the second follows from a relation, 
\begin{equation}
\frac{\partial}{\partial\zeta_{0\mu a}}
(\bar{\sigma}\cdot p-\bar{M})^{-1}
=(\bar{\sigma}\cdot p-\bar{M})^{-1}g\bar{\sigma}^\mu\frac{\rho_a}{2}(\bar{\sigma}\cdot p-\bar{M})^{-1},
\end{equation}
and the last equality follows from 
\begin{equation}
\frac{\partial J^\mu_a}{\partial \zeta_{\nu b}}=m_Y^2g^{\mu\nu}\delta_{ab},
\label{SubC}
\end{equation}
which can be derived from (\ref{SSSBC}).

The logarithmically divergent part of $\Delta{\cal L}(\delta\bm{Y})$ is found exactly proportional to the kinetic term of the SU(2) Yang-Mills gauge field, including the non-Abelian part.
However, we postulated at the beginning that we should omit non-Abelian terms for simplicity. Therefore, in the same level of approximation, we treat $\delta\bm{Y}^{\mu\nu}$ in (\ref{DLFSU(2)}) as not containing a non-Abelian part.
If we had started with local SU(2) symmetry, this approximation would not have been necessary, but in that case, the conservation of SU(2) current discussed in the next section would become somewhat complicated to estimate. The reason for the global SU(2) approximation lies in this point. 
The effect of the non-Abelian part will be of order $(m/m_Y)^2$ to the Abelian part since the vacuum expectation value of the gauge potential is of order $m$. Then, if an emergent fermion mass $m$ is small enough compared with the weak boson mass $m_Y$, the contribution from the non-Abelian part will be negligible. 
The detailed considerations of full gauge symmetry are beyond the scope of this paper.

From (\ref{DLFSU(2)}) with (\ref{SSSBCSU(2)}) and (\ref{Pi^munu_ab}), we have 
\begin{eqnarray}
\Delta{\cal L}(\delta\bm{Y})-\bm{Z}^\mu\cdot\frac{\partial\Delta{\cal L}(\delta\bm{Y})}{\partial\delta\bm{\zeta}^\mu}
&=&-m_Y^2\bm{\zeta}^\mu_0\cdot\delta\bm{\zeta}_\mu\nonumber\\
&&-\frac{m_Y^2}{2}(\delta\bm{\zeta}^\mu\cdot\delta\bm{\zeta}_\mu-\bm{Z}^\mu\cdot\bm{Z}_\mu)
\nonumber\\
&&+\frac{1}{12}g^2K_2(\delta\bm{\zeta}^{\mu\nu}\cdot\delta\bm{\zeta}_{\mu\nu}-\bm{Z}^{\mu\nu}\cdot\bm{Z}_{\mu\nu})
\nonumber\\
&&+\Delta'{\cal L}(\delta\bm{Y})-\bm{Z}^\mu\cdot\frac{\partial\Delta'{\cal L}(\delta\bm{Y})}{\partial\delta\bm{\zeta}^\mu}.
\label{DL=}
\end{eqnarray}
Then, except for a constant term, the effective Lagrangian (\ref{EffecLSU(2)}) becomes 
\begin{equation}
{\cal L}_{\rm eff}=-\frac{1}{4}\bm{Z}_R^{\mu\nu}\cdot\bm{Z}_{R\mu\nu}+\frac{m_Z^2}{2}\bm{Z}_R^\mu\cdot\bm{Z}_{R\mu}
-\frac{1}{4}\delta\bm{\zeta}_R^{\mu\nu}\cdot\delta\bm{\zeta}_{R\mu\nu}
+\varphi'^\dagger\left( \bar{D}-g'\bar{\sigma}^\mu\frac{\bm{\rho}}{2}\cdot\delta\bm{\zeta}_{R\mu}\right)\varphi'-:\bm{j}'^\mu:\cdot\bm{Z}_{R\mu},
\label{EffecLSU(2)RR}
\end{equation}
where $\bm{Z}_R^\mu=\sqrt{1+g^2K_2/3}\bm{Z}^\mu$, $m_Z^2=2m_Y^2(1+g^2K_2/3)^{-1}$, $\bm{\zeta}_R^\mu=\sqrt{-g^2K_2/3}\bm{\zeta}^\mu$, and $g'/g=(-g^2K_2/3)^{-1/2}$. We remember here that $K_2<0$.
In the effective Lagrangian (\ref{EffecLSU(2)RR}), the part $\Delta'{\cal L}(\bm{Y})-\bm{Z}^\mu\cdot\partial\Delta'{\cal L}(\bm{Y})/\partial\delta\bm{\zeta}^\mu$ in (\ref{DL=}) has been replaced by a similar form to that before the path integration of the spinor fields, where the prime symbol on $\varphi$ or $\bm{j}^\mu$ denotes that the divergences arising from loop corrections should be removed by replacing $k_1$ with $5m^2/(72\pi^2)$, and $K_2(m)$ with 0, in accordance with (\ref{Gamma}). 

As expected from Eguchi's paper, the kinetic term for the Nambu-Goldstone photons $\bm{\zeta}_R^\mu$ emerges solely from renormalization. 
We further see in (\ref{EffecLSU(2)RR}) a kind of seesaw mechanism. The fermion loop correction deprives $\delta\bm{\zeta}^\mu$ of its mass term and gives the same amount to $\bm{Z}^\mu$ or, in other words, the $\delta\bm{\zeta}$ bosons become massless while the mass square of the $\bm{Z}$ bosons is doubled. 

In deriving emergent photons, the proof of the vanishing of their mass relies on the relation (\ref{SubC}), which is interpretable as an SU(2) extension of the cancellation mechanism  for the Bjorken-photon mass: Eq. (16) in \cite{Bjorken}, while the cancellation prescription of the Eguchi-photon mass Eq. (3.9) in \cite{Eguchi} is not given in the same form.
It may worth noticing, however, that the explicit evaluation of divergent integral $\Pi^{\mu\nu}_{ab}(0)$ will not confirm 
$\Pi^{\mu\nu}_{ab}(0)=-m_Y^2g^{\mu\nu}\delta_{ab}$ in 
(\ref{Pi^munu_ab}), and therefore, rather, we should postulate it for Lorentz invariance of the emergent theory.
The necessity of this postulation seems to be due to the incompleteness of quantum field theory itself.

Quantum field theory is incomplete in the sense that it contains divergences.
Any regularization scheme will not confirm Ward identities when divergences are explicitly evaluated. Rather, what we can do is to control divergences by using Ward identities. Similarly, divergences in our calculations should be treated so as to make the emergent theory invariant under Lorentz symmetry, which in turn guarantees the vanishing mass of emergent photons in accordance with the Nambu-Goldstone theorem. Consequently, the relation (\ref{Pi^munu_ab}) should  be viewed as an analogue of the Ward identity, applicable to theories with broken Lorentz symmetry.

Besides the global approximation of SU(2) symmetry, the description of the effective theory (\ref{EffecLSU(2)RR}) is not yet complete, since fermions are still described by a chiral doublet, not by the Dirac spinors representing emergent neutrinos and electrons. We have not yet succeeded in obtaining  such a representation of the Lagrangian (\ref{EffecLSU(2)RR}). 

Moreover, we should also mention the difference between the SU(2) symmetry in our chiral model and that in the standard theory. As we have already seen from (\ref{SU(2)equivalence}), $\nu_L$ and $e_L$ are singlets with respect to the original SU(2) in our model, though these emergent fermions should constitute an SU(2) doublet in the sense of the standard theory.
As a result, in order to compare our effective theory with the standard theory, massive vector bosons $\bm{Z}_R^\mu$ and emergent photons $\delta\bm{\zeta}^\mu_R$ should be further rewritten in terms of $W^\mu_\pm$, $Z^\mu$, and $A^\mu$ in the standard SU(2)$\times$U(1) representation. 
Because of this incompleteness, there remain several questions still to be clarified.

One question is as to whether the isotriplet of the Nambu-Goldstone vector bosons is allowable for identifying them with photons in the standard theory, or whether those bosons contain photons in the sense of the standard theory.
Concerning the SU(2) nature of the Nambu-Goldstone photons, it will become understandable if we remember the Nishijima-Gell-Mann formula: $Q=I_3+Y/2$ \cite{NN,Nishijima,GellMann}, in which the isospin takes part of the electric charge. 
On the other hand, since emergent neutrinos and electrons are singlets under the original SU(2) transformations, the action of the photon triplet on emergent leptons will be effectively equivalent to one massless vector field. 
In this sense, only one component in the triplet will correspond to the electromagnetic potential $A^\mu$ in the standard theory. We will discuss in the subsequent sections the roles of the remaining two components of $\delta\bm{\zeta}^\mu_R$ not contributing as the standard photon.

Another question involves how the emergent photons $\bm{\zeta}_R^\mu$ couple with emergent leptons. 
If we consider only the appearance of the effective Lagrangian  (\ref{EffecLSU(2)RR}),  $\bm{\zeta}_R^\mu$ and $\bm{Z}_R^\mu$ would similarly couple with emergent leptons. The differences seem to lie only in the coupling constants and the masses of these bosons.
However, if $\bm{\zeta}_R^\mu$ are to be identified with the photons in the standard theory, they should at least decouple from the emergent neutrinos, while $\bm{Z}_R^\mu$ should not if they can represent neutral currents in the standard theory.

Though the above questions cannot be clarified only from the effective Lagrangian (\ref{EffecLSU(2)RR}),  we present in the next section other arguments which show that the massless triplets $\bm{\zeta}_R^\mu$ decouple from the neutrinos, and part of them contributes to generating a new emergent fermion, which will in turn give rise to asymmetries between $\bm{\zeta}_R^\mu$ and $\bm{Z}_R^\mu$ in interacting with the emergent leptons.
%%%%%%%%%%%%%%%%%%%%%%%%%%%%%%%
\section{Decoupling of neutrinos and emergence of a right-handed electron\label{ERHF}}
In the previous section, we saw that the vanishing of the emergent photon mass is not an automatic consequence of   straightforward calculations, due to divergences inherent in quantum field theory. However, we can avoid this difficulty by considering matrix elements, since matrix elements of some operators, which are directly related to measurements, should be finite, and therefore we can obtain exact results devoid of ambiguities stemming from divergences. 

Actually, the vanishing of mass for the emergent photons can be  confirmed by the appearance of a pole in the matrix elements of the SU(2) current $\bm{j}^\mu$. If its matrix element between one-emergent-lepton states is expressed by
\begin{equation}
\langle p'|j^\mu_a(x)|p\rangle=\varphi_{p'}^\dagger(g\bar{\sigma}^\mu\frac{\rho_a}{2}+\Lambda^\mu_a)\varphi_{p}
\frac{e^{iq\cdot x}}{V},
\end{equation}
the current conservation leads to 
\begin{equation}
q\cdot\Lambda_a=g[\frac{\rho_a}{2}, \bar{M}]
=-\frac{img}{2}(\bm{\rho}\times\bm{\sigma})_a.
\label{LongVertex}
\end{equation}
Then a massless pole appears in the longitudinal part of $\Lambda^\mu_a$:
\begin{equation}
\Lambda^\mu_a({\rm longitudinal})=-\frac{q^\mu}{q^2}\frac{img}{2}(\bm{\rho}\times\bm{\sigma})_a,
\label{MPT}
\end{equation}
which shows that the mass of the emergent photons is exactly zero.
Furthermore, we see that these emergent photons decouple from the neutrinos.
In fact, we can verify by using (\ref{WaveFunctions}) that
\begin{equation}
\langle \nu_{p'}|\bm{\rho}\times\bm{\sigma}|\nu_{p}\rangle=0
\label{DecouplingOfNuFromEP}
\end{equation}
holds, which implies that the pole term in a matrix element for the emergent neutrinos disappears or, equivalently, that the emergent photons do not interact with the neutrinos. 

On the other hand, we see that the emergence of a new fermion state, or a right-handed electron, is demonstrated by the disappearance of the pole term (\ref{MPT}) from the matrix elements for the emergent electrons.
In this case, the divergence of the pole term can be rewritten in the form
\begin{equation}
\varphi_{e\bm{p}'}^\dagger q\cdot\Lambda_a\varphi_{e\bm{p}}=(R\varphi_{e\bm{p}'})^\dagger g\sigma\cdot qR\varphi_{e\bm{p}},
\label{LeftToRight}
\end{equation}
where
\begin{equation}
\begin{array}{cc}
R\varphi_{e\bm{p}}=\lambda_{\bm{p}+}\chi_{\bm{p}L}\varphi_{\bm{p}R}+\lambda_{\bm{p}-}\chi_{\bm{p}R}\varphi_{\bm{p}L},
&\sigma^\mu=(1,\bm{\sigma}),
\end{array}
\end{equation}
which can be verified by the help of the identities obtainable from the conservation of a Dirac current in the 2-spinor representation  given in Appendix \ref{Din2}, for example, 
\begin{equation}
q_\mu\bar{u}_{\bm{p}'L}\gamma^\mu u_{\bm{p}L}
=\lambda_+'\lambda_+\varphi_{\bm{p}'L}^\dagger\bar{\sigma}\cdot q\varphi_{\bm{p}L}+\lambda_-'\lambda_-\varphi_{\bm{p}'L}^\dagger\sigma\cdot q\varphi_{\bm{p}L}=0.
\label{an_identity}
\end{equation}
We can obtain expressions similar to (\ref{LeftToRight}) for the emergent positrons, as well as for the pair creations and annihilations.
As a result, if we define the $R$ operator by
\begin{equation}
\left\{\begin{array}{l}
R\varphi_{\nu\bm{p}}=R\varphi_{\bar{\nu}\bm{p}}=0,\\
R\varphi_{e\bm{p}}=\lambda_{\bm{p}+}\chi_{\bm{p}L}\varphi_{\bm{p}R}+\lambda_{\bm{p}-}\chi_{\bm{p}R}\varphi_{\bm{p}L},\\
R\varphi_{\bar{e}\bm{p}}=\lambda_{\bm{p}+}\chi_{\bm{p}L}\varphi_{\bm{p}R}-\lambda_{\bm{p}-}\chi_{\bm{p}R}\varphi_{\bm{p}L},
\end{array}\right.
\label{R-op}
\end{equation}
then the emergent doublet $R\varphi$ satisfies 
\begin{equation}
[\sigma^\mu i\partial_\mu-\bar{M}]R\varphi=0,
\label{EqMforRphi}
\end{equation}
from which we see that $R\varphi$ is right-handed. 
The massless pole term (\ref{MPT}) in turn transforms to 
\begin{equation}
-\frac{q^\mu}{q^2}\frac{img}{2}(\bm{\rho}\times\bm{\sigma})_a
\simeq R^\dagger g\sigma^\mu\frac{\rho_a}{2}R,
\label{PA}
\end{equation}
in a matrix element for neutral currents.
Comparing (\ref{R-op}) with (\ref{WaveFunctions}), the $R$-operation for the wave functions of emergent electrons is found to be equivalent to the simultaneous spin and isospin reversals.

The equation (\ref{EqMforRphi}) implies that wave function $R\varphi_{e\bm{p}}$ has Hamiltonian $H=\bm{\sigma}\cdot\bm{p}+\bar{M}$, while that for an antiparticle (a hole) $R\varphi_{\bar{e}\bm{p}}$ has $H=\bm{\sigma}\cdot\bm{p}-\bar{M}$. 
Since we find 
\begin{eqnarray}
HR\varphi_{e\bm{p}}=(\omega-\frac{m}{2})R\varphi_{e\bm{p}},&&
HR\varphi_{\bar{e}\bm{p}}=(\omega+\frac{m}{2})R\varphi_{\bar{e}\bm{p}},
\end{eqnarray}
the state $R\varphi_{e\bm{p}}$ has the energy $\omega-m/2$, while $R\varphi_{\bar{e}\bm{p}}$ has the 
energy $\omega+m/2$.
Comparing this with (\ref{DR}), the above considerations verify that the right-handed electron has the same mass $m$ and potential $-m/2$  as the left-handed electron, while the right-handed positron has the same mass $m$ and potential $m/2$ as the left-handed positron, as it should if the emergent theory is Lorentz invariant.

The disappearance of a massless pole from the neutral matrix elements for the SU(2) current is interpretable such that the Nambu-Goldstone vector mesons, or the emergent photons, are absorbed, partly at least, by an emergent electron and endow it with a right-handed state, or create a right-handed isosinglet in the sense of the standard theory.

Viewed from the final situation only, it looks as if the vector mesons transmuted into fermions, which is in contrast to the ordinary Higgs mechanism in which boson states are transformed into other boson states.
The alteration of statistics has been known to occur in an SU(2) model when there exists a 't Hooft--Polyakov monopole \cite{HGT,JR,ASG}. This aspect will be examined in Sec.\ref{FBP}.
%%%%%%%%%%%%%%%%%%%%%%%%%%%%%%%%%%
\section{Emergence of the electromagnetic current}
The emergence of a right-handed electron will also require to modify the isoscalar current:
\begin{equation}
j^\mu_0=\frac{g}{2}\varphi^\dagger\bar{\sigma}^\mu\varphi,
\label{ISC}
\end{equation}
into the effective one,
\begin{equation}
J^\mu_0=\frac{g}{2}\varphi^\dagger(\bar{\sigma}^\mu+R^\dagger \sigma^\mu R)\varphi,
\end{equation}
due to the conservation of the emergent-lepton number.
Combining the emergent-lepton-number current as the zeroth component, we can extend the effective isospin currents to four dimensions:
\begin{equation}
J^\mu_\alpha:=\varphi^\dagger\left[ g\bar{\sigma}^\mu\frac{\rho_\alpha}{2}+R^\dagger g\sigma^\mu\frac{\rho_\alpha}{2}R \right]\varphi, \quad(\alpha=0,\cdots,3).
\label{eSU(2)c4}
\end{equation}
It should be noticed here that the zeroth component of the extended SU(2) current (\ref{eSU(2)c4}) becomes redundant when estimated in a matrix element since it can be reproduced from the isovector part by contracting with an appropriate unit (iso-) vector, for example, $\bm{e}_{\bm{p}}=\bm{p}/\vert\bm{p}\vert$:
\begin{equation}
\chi^\dagger_{s'\bm{p}'}\chi_{s\bm{p}}=s\chi^\dagger_{s'\bm{p}'}\bm{e}_{\bm{p}}\cdot\bm{\rho}\chi_{s\bm{p}},
\end{equation}
where $\chi_{s\bm{p}}$ are isodoublet (\ref{PX}) in the two  component representation, with $s=1$ for the right-handed state, while $s=-1$ for the left-handed state.

The quasi-Higgs mechanism enables us to represent the electromagnetic current $j^\mu_{\rm em}(x)=e\bar{\psi}\gamma^\mu\psi$ in terms of $J^\mu_\alpha$ as follows. 
We first define the coefficients $e_{\alpha s's}$ by 
\begin{equation}
e_{\alpha s's}(\bm{p}',\bm{p}):=\frac{1}{\sqrt{2}}\chi^\dagger_{s'\bm{p}'}\rho_\alpha\chi_{s\bm{p}},
\end{equation}
which satisfy the orthonormal conditions
$$\begin{array}{cc}
\displaystyle\sum_\alpha e_{\alpha s's}e_{\alpha r'r}^\dagger=\delta_{s'r'}\delta_{sr},&
$$\displaystyle\sum_{s's} e_{\alpha s's}e_{\beta s's}^\dagger=\delta_{\alpha\beta}.
\end{array}$$
They transform an isovector to a bispinor. We can actually verify that the equality
\begin{equation}
\langle e',s'|j^\mu_{\rm em}(x)|e,s\rangle=\frac{\sqrt{2}e}{g}\displaystyle\sum_{\alpha=0}^3 c_{\alpha s's}^\dagger\langle e'|J^\mu_\alpha(x)|e\rangle,
\label{J-JR}
\end{equation}
holds, where
\begin{equation}
\begin{array}{cc}
c_{\alpha L'L}=e_{\alpha R'R},&
c_{\alpha L'R}=e_{\alpha R'L},\\
c_{\alpha R'L}=e_{\alpha L'R},&
c_{\alpha R'R}=e_{\alpha L'L}.
\end{array}
\end{equation}
Similar relations are obtainable also for emergent positrons, pair creations and annihilations, provided that
the relations between $c_{\alpha s's}$ and $e_{\alpha s's}$ are properly rearranged.
It is seen from (\ref{J-JR}) that the transformation coefficients $c_{\alpha s's}$ endow emergent electrons in initial and final states, which otherwise have no extra spin degrees of freedom, with two helicity states for each, and reproduce complete spin combinations of amplitudes for the electromagnetic current.

Furthermore, Eq. (\ref{J-JR}) shows that the Lorentz noncovatiant coefficients $c_{\alpha s's}$ cancel the Lorentz noncovatiant isospin dependence of $\langle e'|J^\mu_\alpha(x)|e\rangle$ to produce the ordinary Lorentz covariant amplitudes: $\langle e',s'|j^\mu_{\rm em}(x)|e,s\rangle$.
%%%%%%%%%%%%%%%%%%%%%%%%%%%%%%%%%
\section{Other roles of Nambu-Goldstone bosons\label{OR}}
We here arrive at the point to discuss the question raised at the end of Sec.\ref{EP}.
As we saw in Sec.\ref{ERHF}, one of the roles played by the Nambu-Goldstone bosons which do not emerge as photons is certainly to generate a right-handed state of the electron. However, these two roles do not exhaust all of the Nambu-Goldstone modes. In fact, the quasi-Higgs mechanism does not necessarily remove from amplitudes all of the Nambu-Goldstone vector mesons, since the pole term still remains in matrix elements for charged weak currents:
$\langle e'|j^\mu_a(x)|\nu\rangle$, etc.

It was reported in another paper \cite{KN2} that due to  extra potential terms, our model appears to give for the charged weak interaction processes a different energy conservation from the conventional form. In this respect, the massless pole for charged weak currents may imply that the remaining components of Nambu-Goldstone mesons plays the role of recovering the energy conservation law to the conventional form. In fact, we can see that a massless boson also emerges from the conservation of the energy-momentum tensor as follows.

Lorentz invariance requires the conservation of the symmetric energy-momentum tensor $T^{\mu\nu}$, which in our case is   given by
\begin{equation}
\begin{array}{rl}
T^{\mu\nu}=&\displaystyle\frac{1}{4}\left[\varphi^\dagger(\bar{\sigma}^\mu\nabla^\nu+\bar{\sigma}^\nu\nabla^\mu)\varphi+{\rm H.c.}\right]\\
&+\bm{Y}^{\mu\rho}\cdot\bm{Y}_\rho{}^\nu
+\displaystyle\frac{g^{\mu\nu}}{4}\bm{Y}^{\rho\sigma}\cdot\bm{Y}_{\rho\sigma}\\
&+m_Y^2\left(\bm{Y}^\mu\cdot\bm{Y}^\nu
+\displaystyle\frac{g^{\mu\nu}}{2}\bm{Y}^\rho\cdot\bm{Y}_\rho\right),
\end{array}
\label{SEMT}
\end{equation}
where the covariant derivative $\nabla_\mu$ is defined by $\nabla_\mu=\partial_\mu+i\frac{g}{2}\rho_aY_{\mu a}$.
If we express a matrix element of (\ref{SEMT}) for emergent leptons in the form
\begin{equation}
\frac{1}{4}\varphi_{p'}^\dagger\left[\bar{\sigma}^\mu(p'+p)^\nu+\bar{\sigma}^\nu(p'+p)^\mu+X^{\mu\nu}\right]\varphi_{p}
\frac{e^{iq\cdot x}}{V},
\end{equation}
then the conservation of $T^{\mu\nu}$ gives for the longitudinal part of $X^{\mu\nu}$,
\begin{equation}
 X^{\mu\nu}({\rm longitudinal})=-\frac{p'^2-p^2}{q^2}(q^\mu\bar{\sigma}^\nu+q^\nu\bar{\sigma}^\mu).
\label{IsoScalarPole}
\end{equation}
In contrast to the ordinary Lorentz-invariant theories, this term vanishes neither for neutral nor charged currents since the energies of emergent leptons have, as seen from (\ref{DR}), extra potential terms $\pm m/2$, which suggests the emergence of another Nambu-Goldstone vector meson, and the existence of another Higgs-like mechanism which would absorb this meson to recover the energy conservation law to the conventional form.

It is suggestive to consider that this extra massless vector meson, which would resolve the remarked problem, may be part of the emergent photons. Then, the standard photon, the right-handed electron, and the conservation of effective $T^{\mu\nu}$ will exhaust all of the roles of the three emergent photons that appeared in Sec.\ref{EP}, though confirmation of this remains to be done. 
%%%%%%%%%%%%%%%%%%%%%%%%%%%%%%%%
\section{Fermion-Boson Puzzle\label{FBP}}
The wave function of one-particle state for the emergent neutrinos and electrons  has the Hamiltonian,
\begin{equation}
H=-\bm{\sigma}\cdot\bm{p}+\frac{m}{2}\bm{\rho}\cdot\bm{\sigma},
\end{equation}
from which follows 
\begin{eqnarray}
i\left[H, \bm{T}\right]=0,&&\bm{T}=\bm{x}\times\bm{p}+(\bm{\sigma}+\bm{\rho})/2.
\end{eqnarray}
Here, we find that the isospin contributes to the angular momentum, and the ``total angular momentum" $\bm{T}$ conserves.
A similar incorporation of isospins into spins has been reported to occur in the presence of a 't Hooft--Polyakov monopole \cite{HGT,JR,ASG}, which appears in the system consisting of the SU(2) Yang-Mills fields and the Higgs fields of isotriplet \cite{tHooft,Polyakov}. 
A particle with isospin $\bm{I}$ and spin $\bm{s}$ captured by a monopole conserves the total angular momentum $\bm{J}_\text{tot}=\bm{x}\times\bm{p}+\bm{s}+\bm{I}$. 
The reason for the incorporation of isospins into spins, even in the absence of a monopole is clear since both cases are commonly invoked as a result of the vacuum expectation values of the SU(2) gauge potentials. 

In the presence of monopoles, the ``monopole atom," which consists of a monopole and a charged scalar boson with isospin 1/2, has half-integer angular momentum, and it has been shown that these atoms really obey Fermi staistics \cite{HGT,JR,ASG}. On the other hand, the fermions in our model have 0 or 1 for the magnitude of ``total spin" $\bm{S}=(\bm{\sigma}+\bm{\rho})/2$. Then, if the spin-statistics theorem \cite{Pauli} were applicable to this $\bm{S}$, a question arises as to whether the emergent neutrinos and electrons are really fermions. 
Or, the question can be stated as follows: whether even constant gauge potentials can transmute bosons into fermions, or fermions into bosons. This seems to be a serious question since it is hard to imagine that constant gauge potentials give rise to such a grave consequence as the alteration of statistics.

First of all, we point out that the spin-statistics theorem is based on the causality and the positivity of energy in relativistic quantum theory and, therefore, is not directly applicable  to systems that break Lorentz symmetry.
An emergent theory resulting from spontaneous spacetime symmetry breaking will indeed be relativistic, and it belongs to the category to which the spin-statistics theorem can apply.
However, the Lorentz invariance of the emergent theory is maintained with the help of the contributions from the Nambu-Goldstone bosons, irrespective of whether they can be observed as particles or not.
An effective theory in which the Nambu-Goldstone modes are not taken into account will be equivalent to a system which breaks Lorentz invariance explicitly, where the spin-statistics theorem is not applicable.

Contrastingly, the arguments in previous sections, where the Nambu-Goldstone modes are properly taken into account to preserve Lorentz invariance of the emergent theory, show that though the isospin actually contributes to the spin of emergent fermions through the spin-isospin coefficients $c_{\alpha s's}$  in (\ref{J-JR}), the contribution is not to alter fermions to bosons, but to attribute a new spin state to an emergent electron. This phenomenon is a consequence required to occur due to the Lorentz invariance of the emergent theory.
Therefore, concerning the spontaneous spacetime symmetry breaking in our theory, the question of statistics does not exist. 
The emergence of a right-handed fermion is understandable simply by the absorption of the Nambu-Goldstone bosons by a left-handed fermion, which clearly does not conflict with the spin-statistics theorem. 
Rather, the quantization of $\varphi$ given in (\ref{Operator_varphi}) by the Bose statistics would make the energy spectrum unbounded below, which violates the stability of the vacuum and makes our theory inconsistent.

%%%%%%%%%%%%%%%%%%%%%
\section{Conclusion}
In the standard electroweak theory \cite{Weinberg}, leptons are described by a left-handed doublet and a right-handed singlet for each generation. We have seen that the idea of simultaneous breaking  of gauge and spacetime symmetries reproduces automatically the identical asymmetric fermion configuration from a chiral SU(2) doublet only, which in turn supports the view advocated for in \cite{KN1} that the concept of spontaneous breaking of Lorentz symmetry may be essential for understanding the structure of the standard theory as the laws of the present Universe.

In order to describe massive fermions, the standard theory assumes many Yukawa couplings with Higgs fields.  
Though the  introduction of fermion masses in this way appears to be the only method consistent with the chiral SU(2) symmetry in weak interactions, and experiments also seem to support this description \cite{Boer}, it will still be unsatisfactory from a theoretical point of view since no principle is known to introduce  Yukawa couplings, which also bring into the theory many arbitrary parameters. 

Contrastingly, the quasi-Higgs mechanism found in this paper  requires neither the Yukawa couplings depending on arbitrary parameters nor the preparation of right-handed singlets in advance to distinguish charged leptons from neutrinos. 
Therefore, the spontaneous spacetime symmetry breaking with the quasi-Higgs mechanism provides an alternative possibility for  introducing masses of fermions consistently in the framework of the standard theory with no arbitrary parameters, at least for charged leptons. 

The new mechanism generates the mass of charged lepton by the vacuum expectation values of gauge potentials, not by those of Higgs fields in Yukawa couplings.
Even in this case, however, the Higgs bosons are still indirectly responsible for the masses of charged leptons since the vacuum expectation values of gauge potentials stem from the mass of gauge bosons.

Concerning a triplet of emergent photons, on the other hand,  which are Nambu-Goldstone vector bosons originating from spontaneous Lorentz violation, synthetic considerations based on all of the results obtained in this paper suggest that, after making the emergent theory Lorentz invariant, only one component will effectively remain detectable, which will be identified with the photon in the standard theory.
 
%%%%%%%%%%%%%%%%%%%%%%%%%%%%%%%%%
\begin{acknowledgments}
The author would like to thank F. Englert for his comments on the masses of fermions received at the Yukawa Institute for Theoretical Physics.
\end{acknowledgments}
%%%%%%%%%%%%%%%%%%%%%%%%%%%%%%%%%
\appendix
\section{Lorentz-invariant estimation of $S(0)$\label{LIEDI}}
The Lorentz-invariant estimation of a quadratically divergent integral $S(0)$ in (\ref{DefJ^mu}) requires some caution. 
The $S(0)$ appearing in the definition of $\bm{J}_0^\mu$ (\ref{DefJ^mu}) is rewritten in the form
\begin{eqnarray}
S(0)&=&\displaystyle\int\frac{d^4p}{(2\pi)^4}i
\frac{X_0+\bm{\rho}\cdot\bm{X}}{\vert\Delta\vert},\label{S(0)}\\
X_0&=&\left(p\cdot p-\frac{m^2}{4}\right)\sigma\cdot p-\frac{m^2}{2}\left(p_0-\frac{m}{2}\right),\\
\bm{X}&=&\left(p\cdot p-mp_0+\frac{m^2}{4}\right)\frac{m}{2}\bm{\sigma}-\left(\sigma\cdot p-\frac{m}{2}\right)m\bm{p},\\
\vert\Delta\vert&=&
\left[(p_0-\frac{m}{2})^2-\bm{p}^2+i\epsilon\right]\left[(p_0+\frac{m}{2})^2-\omega^2+i\epsilon\right],
\end{eqnarray}
where $\omega=\sqrt{\bm{p}^2+m^2}$, and the contour of $p_0$ integration has been specified by the infinitesimal imaginary part $i\epsilon$: $(\epsilon>0)$, which is dictated by the hole interpretation of particles corresponding to the negative branch of the square root in the dispersion relation.

We first notice that the iso-scalar part of (\ref{S(0)}) vanishes  since, due to its invariance under space rotations,
\begin{equation}
\int\frac{d^4p}{(2\pi)^4}i
\frac{X_0}{\vert\Delta\vert}=
\int\frac{d^4p}{(2\pi)^4}i\frac{p_0^3-(\bm{p}^2+3m^2/4)p_0+m^3/4}{\vert\Delta\vert},
\label{S(0)_s}
\end{equation}
while the $p_0$ integration gives
\begin{eqnarray}
\int\frac{dp_0}{2\pi}
\frac{i}{\vert\Delta\vert}&=&\frac{1}{4m\bm{p}^2}
\left(-\frac{m}{\omega}\right),\\
\int\frac{dp_0}{2\pi}\frac{ip_0}{\vert\Delta\vert}&=&\frac{1}{4m\bm{p}^2}
\left(\vert\bm{p}\vert-\omega+\frac{m^2}{2\omega}\right),
\\
\int\frac{dp_0}{2\pi}\frac{ip_0^3}{\vert\Delta\vert}&=&\frac{1}{4m\bm{p}^2}
\left(\vert\bm{p}\vert^3-\omega^3+\frac{3m^2}{4}(\vert\bm{p}\vert+\omega)+\frac{m^4}{8\omega}\right),\\
\end{eqnarray}
which makes (\ref{S(0)_s}) vanish.
The isovector part, on the other hand, can be written as 
\begin{equation}
\displaystyle\int\frac{d^4p}{(2\pi)^4}i
\frac{\bm{\rho}\cdot\bm{X}}{\vert\Delta\vert}=
\Gamma\bar{M},
\end{equation}
\begin{equation}
\Gamma=\displaystyle\int\frac{d^4p}{(2\pi)^4}i
\frac{(p_0-m/2)^2-\bm{p}^2/3}{\vert\Delta\vert}
=\displaystyle\int\frac{d^3p}{(2\pi)^3}\frac{1}{3\omega}.
\label{DivGamma}
\end{equation}
In order to separate the divergent part from (\ref{DivGamma}) in a Lorentz-invariant way, we consider a subsidiary integral
\begin{eqnarray}
Z(m^2)&=&\int\frac{d^4p}{(2\pi)^4}i
\frac{\sigma\cdot p\bar{M}\sigma\cdot p}{(p^2+i\epsilon)(p^2-m^2+i\epsilon)}\nonumber\\
&=&\bar{M}\int\frac{d^4p}{(2\pi)^4}i
\frac{p_0^2-\bm{p}^2/3}{(p^2+i\epsilon)(p^2-m^2+i\epsilon)}\nonumber\\
&=&\bar{M}\int\frac{d^3p}{(2\pi)^3}
\frac{1}{3}
\left(\frac{1}{\vert\bm{p}\vert+\omega}+\frac{1}{2\omega}\right),
\end{eqnarray}
while the Lorentz covariant replacement of $p^\mu p^\nu$ with $p^2g^{\mu\nu}/4$ gives $Z(m^2)=K_1\bar{M}/2$.

We next consider another subsidiary quantity $W$:
\begin{eqnarray}
W&=&\frac{4}{3}Z(m^2)-\frac{1}{3}Z(0)\nonumber\\
&=&\left\{
\begin{array}{ll}
\displaystyle\left(\frac{2}{3}K_1-\frac{k_1}{6}\right)\bar{M}
&:\text{covariant},\\  
\displaystyle\int\frac{d^3p}{(2\pi)^3}\frac{1}{9}
\left(\frac{4}{\vert\bm{p}\vert+\omega}+\frac{2}{\omega}-\frac{1}{\vert\bm{p}\vert}\right)\bar{M}&:p_0\text{-integration first}.
\end{array}\right.
\end{eqnarray}
Then, $S(0)-W$ becomes finite:
\begin{eqnarray}
S(0)-W&=&\int\frac{d^3p}{(2\pi)^3}\frac{1}{9}
\left(\frac{1}{\vert\bm{p}\vert}+\frac{1}{\omega}-\frac{4}{\vert\bm{p}\vert+\omega}\right)\bar{M}\nonumber\\
&=&\frac{m^2}{144\pi^2}\bar{M}.
\end{eqnarray}
Accordingly, the Lorentz-invariant estimation leads to  
\begin{eqnarray}
S(0)&=&W+S(0)-W\nonumber\\
&=&\left[\frac{2}{3}K_1-\frac{k_1}{6}+\frac{m^2}{144\pi^2}\right]\bar{M}\nonumber\\
&=&\left[\frac{k_1}{2}+\frac{2m^2}{3}\left(K_2-\frac{5}{96\pi^2}\right)\right]\bar{M},
\label{SWS}
\end{eqnarray}
from which we have (\ref{Gamma}).
The last equality in (\ref{SWS}) follows from the relation $K_1=k_1+m^2(K_2-1/(16\pi^2))$, which can be verified by the following relations:
\begin{eqnarray}
\frac{\partial K_1}{\partial m^2}=K_2,&&
\frac{\partial K_2}{\partial m^2}=2K_3
=\frac{1}{16\pi^2m^2}.
\end{eqnarray}
By integrating the second equation we have
\begin{equation}
K_2=\frac{1}{16\pi^2}\ln\left(\frac{m^2}{\mu^2}\right),
\end{equation}
with some integration constant $\mu$ of mass dimension one.
Then, by integrating the first equation, we find 
\begin{eqnarray}
K_1&=&k_1+\frac{m^2}{16\pi^2}\left(-1+\ln\left(\frac{m^2}{\mu^2}\right)\right)\nonumber\\
&=&k_1+m^2\left(K_2-\frac{1}{16\pi^2}\right),
\end{eqnarray}
where the integration constant has been determined to be 
$k_1$ by taking $m=0$.
%%%%%%%%%%%%%%%%%%%%%%%%%%%%%%%%%%
\section{Dirac spinors in the 2-spinor representation\label{Din2}}
The field operator for a Dirac spinor is conventionally represented in the following form:
\begin{equation}
\psi(x)=V^{-1/2}\sum_{\bm{p},s}\left[a_{\bm{p}s}u_{\bm{p}s} e^{-ip\cdot x}+b_{\bm{p}s}^\dagger v_{\bm{p}s} e^{ip\cdot x}\right],
\end{equation}
where $a_{\bm{p}s}$ is the annihilation operator of a fermion with momentum $\bm{p}$ and spin $s$, while $b_{\bm{p}s}$ is that of an antifermion.
If $\psi(x)$ is expanded in terms of 2-spinors in helicity eigenstates $\varphi_{\bm{p}L}$ and $\varphi_{\bm{p}R}$,
we find that the four component wave functions $u_{\bm{p}s}$ and $v_{\bm{p}s}$ are expressible in the forms
\begin{eqnarray}
u_{\bm{p}L}&=&
\sqrt{\frac{\omega+m}{2\omega}}
\left[\begin{array}{c}
\varphi_{\bm{p}L}\\
\displaystyle\frac{\sigma\cdot\bm{p}}{\omega+m}\varphi_{\bm{p}L}
\end{array}
\right]
=
\frac{1}{\sqrt{2}}
\left[\begin{array}{r}
(\lambda_++\lambda_-)\varphi_{\bm{p}L}\\
-(\lambda_+-\lambda_-)\varphi_{\bm{p}L}
\end{array}
\right],\\
u_{\bm{p}R}&=&
\sqrt{\frac{\omega+m}{2\omega}}
\left[\begin{array}{c}
\varphi_{\bm{p}R}\\
\displaystyle\frac{\sigma\cdot\bm{p}}{\omega+m}\varphi_{\bm{p}R}
\end{array}
\right]
=
\frac{1}{\sqrt{2}}
\left[\begin{array}{r}
(\lambda_++\lambda_-)\varphi_{\bm{p}R}\\
(\lambda_+-\lambda_-)\varphi_{\bm{p}R}
\end{array}
\right],\\
v_{\bm{p}L}&=&
\sqrt{\frac{\omega+m}{2\omega}}
\left[\begin{array}{c}
\displaystyle\frac{\sigma\cdot\bm{p}}{\omega+m}\varphi_{\bm{p}R}\\
\varphi_{\bm{p}R}
\end{array}
\right]
=
\frac{1}{\sqrt{2}}
\left[\begin{array}{r}
(\lambda_+-\lambda_-)\varphi_{\bm{p}R}\\
(\lambda_++\lambda_-)\varphi_{\bm{p}R}
\end{array}
\right],\\
v_{\bm{p}R}&=&
\sqrt{\frac{\omega+m}{2\omega}}
\left[\begin{array}{c}
\displaystyle\frac{\sigma\cdot\bm{p}}{\omega+m}\varphi_{\bm{p}L}\\
\varphi_{\bm{p}L}
\end{array}
\right]
=
\frac{1}{\sqrt{2}}
\left[\begin{array}{r}
-(\lambda_+-\lambda_-)\varphi_{\bm{p}L}\\
(\lambda_++\lambda_-)\varphi_{\bm{p}L}
\end{array}
\right],
\end{eqnarray}
which satisfy $u_{\bm{p}s}^\dagger u_{\bm{p}s'}=v_{\bm{p}s}^\dagger v_{\bm{p}s'}=\delta_{ss'}$.
In the above representation, a matrix element of Dirac current $\bar{u}_{\bm{p}'L}\gamma^\mu u_{\bm{p}L}$, for example, is expressible as 
\begin{equation}
\bar{u}_{\bm{p}'L}\gamma^\mu u_{\bm{p}L}
=\lambda_+'\lambda_+\varphi_{\bm{p}'L}^\dagger\bar{\sigma}^\mu\varphi_{\bm{p}L}+\lambda_-'\lambda_-\varphi_{\bm{p}'L}^\dagger\sigma^\mu\varphi_{\bm{p}L}.
\end{equation}
Then we can easily obtain the identities necessary to prove (\ref{LeftToRight}) in Sec. \ref{ERHF} by using the formula
\begin{equation}
\begin{array}{cc}
\left\{
\begin{array}{c}
\bar{\sigma}\cdot p\varphi_{\bm{p}L}=(\omega-\vert\bm{p}\vert)\varphi_{\bm{p}L}=m\displaystyle\frac{\lambda_-}{\lambda_+}\varphi_{\bm{p}L},\\
\sigma\cdot p\varphi_{\bm{p}L}=(\omega+\vert\bm{p}\vert)\varphi_{\bm{p}L}=m\displaystyle\frac{\lambda_+}{\lambda_-}\varphi_{\bm{p}L},
\end{array}\right.
&
\left\{
\begin{array}{c}
\bar{\sigma}\cdot p\varphi_{\bm{p}R}=(\omega+\vert\bm{p}\vert)\varphi_{\bm{p}R}=m\displaystyle\frac{\lambda_+}{\lambda_-}\varphi_{\bm{p}R},\\
\sigma\cdot p\varphi_{\bm{p}R}=(\omega-\vert\bm{p}\vert)\varphi_{\bm{p}R}=m\displaystyle\frac{\lambda_-}{\lambda_+}\varphi_{\bm{p}R}.
\end{array}\right.
\end{array}
\end{equation}

\end{document}